\documentstyle[12pt]{article}

\newcommand{\be}{\begin{equation}}
\newcommand{\ee}{\end{equation}}
\newcommand{\bea}{\begin{eqnarray}}
\newcommand{\eea}{\end{eqnarray}}

\begin{document}
\begin{titlepage}

\begin{flushright}
{\today}
\end{flushright}
\vspace{1in}

\begin{center}
\Large
{\bf High Energy  Scatterings in Higher  Dimensional Theories   }
\end{center}

\vspace{.2in}

\normalsize

\begin{center}
{ Jnanadeva Maharana \footnote{In memory of my Sumitra } \\
E-mail maharana$@$iopb.res.in} 
\end{center}

\normalsize

\begin{center}
 {\em Institute of Physics \\
Bhubaneswar - 751005, India  \\
and \\
Max-Planck-Institut f\"ur Gravitationsphysik (Albert-Einstein-Institut)    ,\\
DE-14476 Potsdam, Germany    }

\end{center}

\vspace{.2in}

\baselineskip=24pt

\begin{abstract}
The high energy behavior of scattering amplitudes in
spacetime dimensions, $ D > 4$, is investigated.
The bound on total cross sections,
 $\sigma_t \le Constant~ (log s)^{D-2}$, $D\ge 4$ has been obtained in the past
 under usual assumptions. I derive new bounds on scattering amplitudes in the
 region $|t|< T_0$,  $t$ being momentum transfer squared
and $T_0$ is a constant.
The existence of a zero-free region for the amplitude,
in complex t-plane, is proved.
I prove stronger upper and lower bounds for the absorptive amplitude in the
domain $0<t<T_0$.
\end{abstract}

\vspace{.5in}

\end{titlepage}




\bigskip
The Froissart-Martin bound restricts the growth of
hadronic total cross sections, $\sigma_t(s)$, at high energies.
 Heisenberg \cite{w}, in 1952,
argued that total cross sections may grow as fast as
$(ln s)^2$ at high energies.
The arguments were based on his intuitions and  he supported them by deriving
the
energy dependence of $\sigma_t$ from the Born-Infeld action for
a scalar field. \\
Let us summarize
the essential  ingredients for derivation of rigorous results in strong
interactions. These results are especially important in the 
context of  hadronic reactions at asymptotic energies.
We refer to books and reviews
\cite{book1,eden,roy} for detail discussions and
references. We consider scattering of scalar particles though out.
The  derivation of
 exact results  rest on four pillars based on axiomatic field theory:\\
(A1) Unitarity of S-matrix.\\
(A2) The amplitude, for fixed $s$, is analytic
in
complex $cos~\theta$-plane inside the Lehmann-Martin ellipse; $s$ and $\theta$
being the energy squared and scattering angles, respectively, in the center of
mass frame.
 The foci of the ellipse lie at $(-1,+1)$ and its
semi-major axis is $cos~\theta_0=1+2t_0/s$, $t_0$ being an $s$-independent
constant. For many hadronic processes, $t_0=4m_{\pi}^2$. The partial wave
expansion of the
amplitude, $F(s,t)$, converges absolutely inside Lehmann-Martin ellipse,
\be
\label{lexpand}
 F(s,t)={{\sqrt s}\over k}\sum _{l=0}^{\infty}(2l+1)f_l(s)P_l(cos\theta)
\ee
$P_l(cos\theta)$ is analytic
in $cos\theta$, so is $F(s,t)$  in $t$, $cos\theta=1+2t/s$; $k$ being the
$c.m.$ momentum. Unitarity  bounds  on
the partial-wave amplitudes are
\be
\label{unitaryf}
0\le |f_l(s)|^2\le {\rm Im}f_l(s)\le 1
\ee
(A3). Polynomial boundedness inside Lehmann-Martin ellipse \cite{egm}.
For $0\le t\le t_0$
\be
\label{polbound}
|F(s,t)|<_{|s|\rightarrow\infty}~ |s|^N
\ee
N being a positive integer.\\
(A4). $F(s,t)$ is analytic in  the complex $s$-plane. There are cuts in the
s-plane as a consequence of  $s$-channel unitarity and crossing symmetry -
crossing is  a requirement. The Froissart-Martin bound is \cite{fr,martin}
\be
\label{m-fbound}
\sigma_t(s)<{{4\pi}\over{t_0-\epsilon}}[ln(s/s_0)]^2
\ee
$\epsilon$ is an arbitrary small constant and $s_0$ is an undetermined energy
scale. However, recently it has been shown \cite{mr}  that
${s_0}^{-1} =17\pi{\sqrt{\pi/2}}{m_{\pi}}^{-2}$ for $\pi\pi$ scattering.
There are host
of important, exact and well tested
bounds on \cite{book1,eden,roy}
\be
\label{allbounds}
\sigma_t,~\sigma_{el},~({{d\sigma}\over{dt}})_{el},~b(s),~
{\bf h}_d={{d\sigma}\over{dt}}|_{t=0}
\ee
where $b(s)$ is the
 slope of the elastic diffraction peak and ${\bf h}_b$ is its height at $t=0$.
  \\
The purpose of this article is to derive some important
bounds for amplitudes describing scattering in $D>4$.
The bounds on $\sigma_t$ and slope of the diffraction peak
associated with the imaginary part of the amplitude etc.
 have been derived in the past \cite{moshe1,moshe2}.  I derive new bounds on
the elastic amplitude
$F^{\lambda}(s,t)$, defined below, in a region $|t|\le T_0$,
$T_0$ being a constant. This result extends the previously obtained results of
\cite{moshe1,moshe2} to a region alluded to above.
Moreover, a bound is derived on the distribution of
zeros of
$F^{\lambda}(s,t)$
in the  complex $t$-plane. The radius of the  zero-free circle
shrinks as $(1/{lns})^2$. I prove that
the absorptive amplitude is bounded from above and below in the
domain $0<t<T_0$.
There is a close relationship
between the energy dependence of $b(s)$  and the zeros of $F^{\lambda}(s,t)$.
The assumptions leading to these bounds are
stated in sequel;
although these are  quite
general, they do not follow from   axiomatic field theory like
(A1)-(A4) for $D=4$ case. 
\\
We recall that, for $D>4$,  the elastic scattering amplitude of spinless
particles admits  partial wave expansion as demonstrated by
Soldate \cite{soldate}.
The  amplitude depends on $s$ and
$cos\theta$. The ultraspherical Jacobi function,
the Gegenbauer polynomials \cite{szego,bateman},  are the basis as
these are the
 'spherical harmonics' associated with the $SO(D-1)$
rotation group.
 The
 amplitude, $F^{\lambda}(s,t)$, is expanded as
\be
\label{gexpand}
F^{\lambda}(s,t)=A_1s^{-\lambda+1/2}\sum_{l=0}^{\infty}(l+\lambda)f^{\lambda}_l
(s)C_l^{\lambda}(t)(1+2t/s)
\ee
where $\lambda={1\over 2}(D-3)$ and    $(s,t)$ are the usual
Mandelstam variables.
$A_1= 2^{4\lambda+3}\pi^{\lambda}\Gamma(\lambda)$, independent of $s$ and $t$.
$C_l^{\lambda}(x)$ are Gegenbauer polynomials satisfying orthogonality
conditions with weight factor $(1-x^2)^{{D\over 2}-1}$, $-1\le x\le+1$
\cite{szego}.
The expansion (\ref{gexpand}) converges in the domain
$-1\le cos\theta \le +1$ \cite{szego}. The partial-wave amplitudes,
 $\{ f_l^{\lambda}(s) \}$, satisfy the
unitarity constraints:
\be
0\le |f^{\lambda}_l(s)|^2 \le {\rm Im}f^{\lambda}_l(s)\le 1
\ee
as derived in \cite{soldate}.\\
Additional assumptions are needed  to derive analog of
Froissart-Martin bound (\ref{m-fbound}).
The existence of extended domain of
analyticity for (\ref{gexpand}) is not proven for the theories in $D>4$, unlike
the $D=4$ case.
Thus two reasonable requirements were imposed in \cite{moshe1,moshe2}. These
are: (I)
The domain of convergence, $\bf D$, of $F^{\lambda}(s,t)$ is an extended
ellipse with  the semimajor axis
 $1+2T_0/s$ in the $t$-plane. (II) The amplitude
is polynomially bounded: $|F^{\lambda}(s,t)|<Cs^N$ in $\bf D$; C and N are
undetermined constants,
 $N\in {\bf R}$.
The domain of convergence of Gegenbauer polynomial
is $[-1,+1]$ and a theorem (see Theorem 9.1.1 of \cite{szego})
 states the convergence properties of  functions  such as $F^{\lambda}(s,t)$.
The partial wave amplitudes
$f_l^{\lambda}(s)$ decay exponentially with $l$ \cite{moshe1,moshe2}
under the  assumptions (I) and (II).
Therefore, the
expansion (\ref{gexpand}) can be truncated at
$L={1\over 2}(N-1){\sqrt{s/{T_0}}}ln s$ and importantly,
 $L$ is $D$-independent. Consequently,  $\sigma_t$
 is bounded \cite{moshe1,moshe2} as
\be
\label{d-bound}
\sigma_t\le C_0(lns)^{D-2}
\ee
Note that $lns$ is to be understood as $ln({s/{\hat s}})$, $\hat s$ is
like $s_0$ which scales $s$. Important bounds on $F^{\lambda}(s,t)$ and its
absorptive part are derived by utilizing crucial properties of
$C_l^{\lambda}(x)$ and the assumptions  (I) and (II).
Let us summarize the useful inequalities
as \cite{szego}.\\
{\it Lemma 1.} For $1 <x_1<x_2< 1+T_0/s$, $C_l^{\lambda}(x_1)<C_l^{\lambda}(x_2)$.
\\
Proof. Define $z={1\over 2}(1-x)$. For $x>1$, $z<0$. $C_l^{\lambda}(x)$, as a
hypergeometric function is expanded as \cite{bateman}
\be
\label{z-exp}
C_l^{\lambda}(x)={{\Gamma({{\lambda+1}\over 2})}\over{\Gamma(2\lambda)}}
\sum_{k=0}^l{{\Gamma(2\lambda+n+k)}\over{k!(l-k)!
\Gamma(\lambda+k+1/2)}} (-z)^k
\ee
 Notice: (i) Coefficients of $(-z)^k$ are positive. (ii) For $1<x_1<x_2$,
 each term in the expansion (\ref{z-exp}), $B_k(z_1)< B_k(z_2)$ and therefore,
$C_l^{\lambda}(x_1)<C_l^{\lambda}(x_2)$ in the range of interest.
Moreover, $C_l^{\lambda}(x)$ oscillates  in the
physical region, $-1\le x< +1$, $x=cos\theta$,  $t<0$.
We  derive following bound on the modulus of the amplitude,
 $|F^{\lambda}(s,t)|$,
  $0<|t|<T_0$, utilizing the Lemma 1 and noting that
$C_l^{\lambda}(1+2t/s)\le C_l^{\lambda}(1+2|t|/s) $.                               \\
{\it Theorem}. The modulus $|F^{\lambda}(s,t)|$ is bounded as
\be
\label{bound1}
|F^{\lambda}|\le A_2({1\over |t|})^{{\lambda+1}\over 2}
({1\over{T_0}})^{\lambda/2}s^{1+(N-1){\sqrt{{|t|\over {T_0}}}}} (lns)^{\lambda}
\ee
$0<|t| <T_0$ and
 $A_2=2\lambda A_12^{-2\lambda-1}\Gamma^{-2}(\lambda+1)$.\\
{\it Proof}. The proof proceeds  through the  following steps.\\
Decompose $F^{\lambda}(s,t)$ as
\be
 F^{\lambda}(s,t)=F^{(1)\lambda}(s,t)+F^{(2)\lambda}(s,t)
 \ee
The expansion of $F^{(1)}(s,t)$ is terminated at $l=L$ and expansion of
$F^{(2)}$
is from $L+1$ to $\infty$. Therefore,
\bea
\label{f1}
F^{(1)\lambda}(s,t)\le A_1s^{-\lambda+1/2}\sum_{l=0}^L(l+\lambda)|f_l^{\lambda}|
C_l^{\lambda}(1+{{2|t|}\over s})
\eea
 Invoking unitarity and setting
 $|f_l^{\lambda}(s)|=1$;   (\ref{f1}) reduces to a sum:
  $ \sum (l+\lambda)C_l^{\lambda}(1+2|t|/s)$. For large $s$,
 \be
\sum_{l=0}^L(l+\lambda)C_l^{\lambda}(1+2|t|/s)\sim C_1
({{\sqrt{s\over |t|}}})^{{1+\lambda}\over 2}L^{\lambda}{\rm exp}
(2L{\sqrt{s\over |t|}})
\ee
Note $L= {1\over 2}(N-1) {\sqrt{s\over{T_0}}}ln s$ and
$C_1=2\lambda 2^{-2(\lambda+1)}\Gamma(\lambda+1)$. Thus the desired bound is
\bea
\label{f1bound}
|F^{(1)\lambda}(s,t)|\le &&A_2({{N-1}\over 2})^{\lambda}
({1\over{|t|}})^{{1+\lambda}\over 2}({1\over{T_0}})^{\lambda/2}\nonumber\\&&
s(lns)^{\lambda}
{\rm exp}{(2L{{\sqrt{{|t|}\over s}}})}
\eea
with $A_2=A_1C_1$. Let us
turn the attention on $F^{(2)\lambda}$.  We argue,
as before, to convince the reader that
\be
\label{f2}
|F^{(2)\lambda}(s,t)|\le A_1s^{-\lambda+1/2}\sum_{L+1}^{\infty}(l+\lambda)
|f_l^{\lambda}|(s)C_l^{\lambda}(1+2|t|/s)
\ee  
and re-express the right hand side (\ref{f2}) as
\bea
\label{schwarzin}
|F^{(2)\lambda}(s,t)|\le &&A_1s^{-\lambda+1/2}\sum_{L=1}^{\infty}\bigg[{\sqrt{
(l+\lambda)}{{C_l^{\lambda}(1+2|t|/s)}\over{{\sqrt{C_l^{\lambda}(1+2T_1/s)}}}}}
\bigg]\nonumber\\&&
\times \bigg[{\sqrt{(l+\lambda)}}{\sqrt{C_l^{\lambda}(1+2T_1/s)}}
|f_l^{\lambda}(s)| \bigg]
\eea
where $|t|<T_1=T_0-\delta_1$;  $\delta_1>0$ but infinitesimally small.
Now, invoke the Schwarz inequality on the {\it r.h.s} of (\ref{schwarzin})
\be
F^{(2)\lambda}(s,t)\le A_1s^{-\lambda+1/2}{\cal S}_1{\cal S}_2
\ee
where
\be
{\cal S}_1= \bigg[\sum_{L+1}^{\infty}(l+\lambda)
{{\bigg(C_l^{\lambda}(1+2|t|/s)\bigg)^2}\over{C_l^{\lambda}(1+2T_1/s)}}
\bigg]^{1/2}
\ee
\be{\cal S}_2=\bigg[\sum_{L+1}^{\infty}(l+\lambda)|f_l^{\lambda}|^2
C_l^{\lambda}(1+2T_1/s) \bigg]^{1/2}
\ee
In what follows, I  outline the estimation of the upper bounds to
$F^{(2)\lambda}(s,t)$ which
is a generalization of the proof
for the case  $ \lambda=1/2$.
The term ${\cal S}_1$  involves ratio of two
 Gegenbauer polynomials with arguments $(1+2|t|/s)$ and $(1+2T_1/s)$. In the
large $s$, limit
${\cal S}_1 \sim (lns)^{\lambda -1}$. For  ${\cal S}_2$, invoke
unitarity, $|f^{\lambda}_l|^2\le1$, and polynomial boundedness, (recall
$|F^{(2){\lambda}}(s,t)|<Cs^N$), to give  suppressed the energy dependence with
appropriate choice of C (see \cite{roy} for detail arguments).
  Thus
 $|F^{(2)\lambda}|$  has subdominant energy dependence:
$(lns)^{\lambda -1}$ ( compared to   $|F^{(1)\lambda}| \sim
(lns)^{\lambda} $).
Therefore,
\be
\label{proof}
|F^{\lambda}(s,t)|\le A_2({1\over{|t|}})^{{1+\lambda}\over 2}
({1\over{T_0}})^{\lambda/2}s^{1+(N-1){{\sqrt{|t|\over{T_0}}}}} (lns)^{\lambda}
\ee
{\it Remarks:} (R1.1) The bound holds in the domain $|t|<T_0$, including a part
  of the physical region  $ t<0$. (R1.2) This is generalization of the bound
for the $D=4$; note the power dependence  and  on $|t|$. The known
result \cite{roy}
 is recovered, setting  $\lambda=1/2$. (R1.3) An important and noteworthy
point is
 the appearance of Regge-like power law in  (\ref{proof}), in the domain
$0<|t|<T_0$ and the  power of
$s$ is $\lambda$-independent. It is a reminiscence
of open string tree amplitude in the Regge limit.
 (R1.4) The bound is a substantial  improvement over that of
\cite{moshe1,moshe2} since theirs  was only for $|F^{\lambda}(s,0)|$.\\
A  bound on number of zeros of $F^{\lambda}(s,t)$ follows from (\ref{proof}).\\
{\it Lemma 2.} The number, $n_r(s)$, of zeros of $F^{\lambda}(s,t)$ within the
disk $|t|<r<T_0$ is bounded from above by
\be
\label{zeros}
n_r(s)\le{{e{\sqrt r}\over{2\sqrt {T_0}}}}lns
\ee
The terms of $o((lns)^{-1})$ in the $r.h.s$ of   (\ref{zeros}) are ignored.\\
{\it Proof}. According to Jensen's theorem \cite{tit}, $n_r(s)$ satisfies the
inequality
\be
\label{jensen}
n_r(s)\le {{1\over{ln{1/{\delta}}}}}{\rm Max}~ln|{{F^{\lambda}(s,r/{\delta})}
\over{F^{\lambda}(s,0)}}|,~~~~r/{T_0}<\delta<1
\ee
Note that $|F^{\lambda}(s,0)|={\sqrt{(Re F^{\lambda}(s,0))^2+(ImF^{\lambda}
(s,0))^2}}$ and $|F^{\lambda}(s,0)|\ge s\sigma_t$ as per our normalization:
$Im F^{\lambda}(s,0)=s\sigma_t$. Therefore,
\be
n_r(s)\le     {{1\over{ln{1/{\delta}}}}} {\rm Max}\bigg(
F^{\lambda}(s,{r\over{\delta}})
{1\over{s\sigma_t}} \bigg)
\ee
by the virtue of (\ref{proof}), on the $r.h.s$ we are
left with
\be
ln \bigg[C_2{\rm exp}(\{(N-1){{\sqrt{{r/{\delta}}\over{T_0}}}}lns\})
({1\over{r/{\delta}}}
)^{{1+\lambda}\over 2} ({{1\over {T_0}}})^{\lambda/2} {{1\over{\sigma_t}}}\bigg]
\ee
Retaining  the leading order term in $lns$
\bea
\label{lnseff}
ln|{{F^{\lambda}(s,r/{\delta})}\over{F^{\lambda}(s,o)}}|&&\le
lns\bigg[ (N-1) {{\sqrt{r\over{\delta}}}}+
{{1\over{(lns)}}}\{lnC_1 +\nonumber\\&&\lambda lnlns\nonumber -
({{{(\lambda+1)\over 2}}}){r\over{\delta}}
- {\lambda\over 2}T_0-ln(\sigma_t)\}\bigg]
\eea
No rigorous lower bound exists on $\sigma_t $ for $D>4$ like the
Jin-Martin \cite{jml} bound: $\sigma_t\ge s^{-6}$ in
 $D=4$. Even if it existed,
 its contribution to (\ref{lnseff}) will be
$(lns)^{-1}ln\sigma_t \sim o(1)$.
Thus retaining the $lns$ term, in large $s$ limit, is justified
in  estimation of the upper bound on $ln|F^{\lambda}(s,r/{\delta})/F^{\lambda}
(s,0)|$; consequently,
\be
\label{jensen1}
n_r(s)\le {{(N-1){\sqrt{{r\over\delta}}}\over{{(ln{1/\delta})}{\sqrt{T_0}}}}}
lns
\ee
follows from (\ref{jensen}). Optimizing
(\ref{jensen1}) with respect to $\delta$ leads to  the desired bound
(\ref{zeros}).\\
{\it Corollary}: There are no zeros inside a disk of
radius $r_0$ in the $t$-plane
and
\be
\label{zero-free}
r_0<{{C_2}\over{(lns)^2}}
\ee
{\it Remarks:} (R2.1) $C_2= T_0[e(N-1)]^{-2}$, is independent of $s$.
The correction to (\ref{jensen1}) is order $o(1)$ if Jin-Martin exists
 for $D>4$.
 (R2.2) There is a zero-free disk with shrinking radius.
More importantly, to
leading order in $lns$, the upper bound on $r_0$ is $D$-independent. (R2.3)
There is a single zero in the  $|t|$-plane in an annular  region:
\be
\label{one-zero}
r_0(s)<{{C_2}\over{(lns)^2}}<r_1(s)<{{C_3}\over{(lns)^2}}
\ee
$C_3$ being another  s-independent constant. We recall
 the study of the zeros of amplitudes has
played  a crucial role in deriving many rigorous results
in the past \cite{bessis,akm,ek}.
\\
The bound on $A^{\lambda}(s,t)= {\rm Im}~F^{\lambda}(s,t)$ can be improved
substantially in the unphysical region $0<t\le T_0$, utilizing unitarity,
properties of Gegenbauer polynomials and the bound (\ref{proof}).
The important point to note is
that $A^{\lambda}(s,t)>0$ for $t>0$ as  evident from (\ref{gexpand}) since
$0\le {\rm Im~f}^{\lambda}_l\le 1$ and $C^{\lambda}_l(x)>0$ for $x>1$. We
arrive at
the final result through following steps.
 The zeros of Jacobi
polynomial $P^{(\alpha,\beta)}_{l}(cos\theta)$, for $\alpha>-1,~\beta>-1$
satisfy following properties: (
Theorem 8.9.1 of \cite{szego}), let
$0<\theta_1<\theta_2,\ldots,<\theta_l<\pi$
be zeros of $P^{(\alpha,\beta)}_{l}(cos\theta)$ (it applies to
$C^{\lambda}_l(cos\theta)~{\rm  since}~\alpha=\beta=\lambda> 1/2 $). Then
\be
\label{zerospace}
\theta_{\nu}={1\over l}(\nu\pi+O(1))
\ee
with $O(1)$ being uniformly bounded constant for all values of
$\nu=1,2,\ldots,l;
~l=1,2,\ldots$. Thus two consecutive zeros are separated \cite{szego} as
$\theta_{\nu+1}-\theta_{\nu}={{\pi}\over l}$. The distribution of zeros of
Legendre polynomial (a special case of Gegenbauer polynomial with $\lambda=1/2$)
was crucially used by Kinoshita to derive an important bound \cite{tk66} on
absorptive amplitude, for $D=4$, in the region $0<t<t_0$. 
We recall that the Gegenbauer polynomials are the ultraspherical Jacobi
polynomials and the distribution of the zeros of the Gegebauer polynomials are
governed by the theorem stated above. Therefore, for the case at
hand, the spacing of zeros (\ref{zerospace}) is also quite analogous. Factorize,
$C^{\lambda}_l(z)={\bf\Pi}_{\nu=1}^l[{{(z-z_{\nu})}\over{1-z_{\nu}}}]$,
$z_{\nu}=cos{\theta_{\nu}}$ are the location of the zeros.
 $A^{\lambda}(s,t)$
becomes a complex function when $t$ is complex i.e. when
$C^{\lambda}_l(z)$ becomes complex. Thus if ${\bf \Phi}$ is the  phase
of $C^{\lambda}_l(z)$ for $z=1+a+ib$, we look for the curve in $z$-plane
which intersects the {\it ellipse} where $C^{\lambda}_l(z)$  hits its first
zero i.e. ${\bf \Phi}={\pi}/2$.
For  complex $t$, decompose
$t=u+iv$, $u$ and $v$ being the real and imaginary parts of $t$.
There is a small domain, $\cal D$, in the upper half $t$ plane, which
is intersection of this curve with  the {\it ellipse} such
that ${\rm Re}~A^{\lambda}(s,t)$ is positive inside ${\cal D}$.
Recall that as we go off real $t$ axis into this domain $A^{\lambda}(s,t)$
 becomes
complex. The arguments of \cite{tk66} can be adopted to show that $\cal D$
is given by
\be
|v|\le{ {\pi{\sqrt u}}\over {2C_4ln s}}
\ee
$C_4$ is $s$-independent constant. Thus ${\rm Re} ~A^{\lambda}(s,t)$ is a
 positive, harmonic function inside ${\cal D}$.
Consequently, a powerful theorem on positive
harmonic functions holds \cite{car}. Now define
$R_0, ~0<R_0<T_0-\delta$, such that it is within $\cal D$. Consider a
disk $|t-R_0 <\pi ({\sqrt{R_0}})/(2C_4lns)$ which is shrinking as $ln s$,
note that $t>0$.\\
The {\it Harnack's theorem} \cite{car} states: For any $t$ in the smaller disk
\be
\label{theo1}
|t-R_0|<{{\pi r{\sqrt{R_0}}}\over{2C_4ln s}}, ~~~0<r<1
\ee
with $r$ inside the disk, the positive, harmonic function,
${\rm Re}~A^{\lambda}(s,t)$ is bounded from above and
below as
\be
\label{theo2}
{{1-r}\over {1+r}}A^{\lambda}(s,R_0)<{\rm Re}~A^{\lambda}(s,t)<{{1+r}\over{1-r}}
A^{\lambda}(s, R_0)
\ee
$A^{\lambda}(s,R_0)$ is real and positive as argued before.
$A^{\lambda}(s,t)$ is defined inside the circle which lies in $\cal D$ and
hence ${\rm Re}~A^{\lambda}(s,t)>0$. Note that the radius is order
$  {1\over{lns}}$, therefore, inequality (\ref{theo2}) conveys that
${\rm Re}~A^{\lambda}(s,t)$ will not increase by more than
a finite factor when $t$ is
increased by ${1\over{lns}}$.
We can derive various bounds on $A^{\lambda}(s,t)$ and its derivatives through
applications of theorem (\ref{theo2}) inside the disk (\ref{theo1}). \\
An upper bound on
$b(s)$
is derived
 utilizing the bound (\ref{proof}) and  the inequality $|F^{\lambda}(s,0)|\ge
s\sigma_t$. It follows from  Cauchy's
inequality and suitable adaptation of  the
arguments of \cite{ek,m1}.
 If the zero-free radius shrinks less rapidly than (\ref{zero-free}) i.e.
$r_0(s)>(lns)^{-2+\eta},~\eta>0$, then bounds on $b(s)$ and $\sigma_t$
are improved \cite{m1}.
 \\
We conclude this note with following remarks and observations in the context 
of higher dimensional theories. There are scenarios where 
 higher dimensional theories  admit  a low compactification
scale, $\sqrt {\hat s}\sim $  a few TeV.
The phenomenological consequences of such models
have been explored extensively \cite{anto,anto1}.
Petrov \cite{petrov} has argued that the cross section of a $D>4$ theory
will have same behavior as that of a $D=4$ theory below the
compactification scale. He supported this claim through 
 model calculations. We mention in passing that 
 the S-matrix has interesting analyticity properties in the energy plane in
case of potential scattering when some spatial dimensions are compactified
\cite{khuri,andre}. These issues have not been thoroughly investigated
in field theoretic frame work.\\
Let us consider a higher dimensional theory with a low energy scale
of compactification, $\hat s$. 
When the theory is probed with energies below the
compactification scale it will behave like a $D=4$ theory as argued
by Petrov. 
However, above  the scale $\hat s$, $\sigma_t$ of a $D>4$ theory
 might behave  as if
(\ref{m-fbound}) is violated, although $\sigma_t$  does {\it not necessarily
have to violate} $D=4$ bound (\ref{m-fbound}).
Thus
it might be  worth while to explore qualitatively whether $\sigma_t$ admits an
$s$-dependence: $(ln{s\over{\hat s}})^{\beta},~\beta>2$ at the extremely high
energies  accessed by LHC and in cosmic ray experiements. If the total cross
section data for extreme high energies show such an energy dependence then 
one might get a hint of decompactification at lower scale as alluded to 
earlier. Moreover,
 there are other  avenues to test  unitarity
bounds on quantities listed in (\ref{allbounds}) for $D=4$.
They
will be  measured with accuracy  at LHC from $s=36~{\rm TeV}^2$ to
 $s=196$ ${\rm TeV}^2$ at LHC.  Moreover,
 $b(s)$ and ${\bf h}_d$ are bounded
by $(lns)^2$; as   ${\bf h}_d$  grows
with energy, the width of ${{d\sigma}\over {dt}}$ shrinks \cite{roy,ek}. Thus
 precision  measurements  of $b(s)$, ${\bf h}_d$ and other
measurable quantities listed in
(\ref{allbounds}) will stringently test unitarity constraints at LHC.
An  unambiguous
 deviation from unitarity  bounds, derived for $D=4$ theories,
 might provide an indirect evidence for existence of higher dimensional
theories.    
\\
Acknowledgments: I am grateful to
I. Antoniadis for valuable suggestions. I thank  S. B. Giddings and
 R. Sasaki for useful correspondence.
I am thankful to Sanjay  Swain for suggestions
 and for his critical remarks.
The warm hospitality  Hermann Nicolai
and the Albert Einstein Institute is gratefully acknowledged.

\end{document}